\documentclass[final, twocolumn,amssymb]{revtex4}

\usepackage{amsmath} 
\usepackage{amssymb} 
\usepackage{longtable}
\usepackage{epsfig} \usepackage{bm}
\usepackage{ulem}

\def \To{{T_\mathrm{o}}}

\begin{document} 
\title{Corresponding States of Structural Glass Formers. II}
\author{Yael S. Elmatad}
\affiliation{Department of Chemistry, University of California, Berkeley, CA 94720, USA}
\author{David Chandler}
\email{chandler@cchem.berkeley.edu}
\affiliation{Department of Chemistry, University of California, Berkeley, CA 94720, USA}
\author{Juan P. Garrahan}
\affiliation{Department of Physics and Astronomy, University of Nottingham, Nottingham, NG7 2RD, United Kingdom}
\date{\today}

\begin{abstract}
The earlier paper of this same title demonstrated a collapse of relaxation data of fragile supercooled glass forming liquids 
[\textit{J. Phys. Chem. B} \textbf{113}, 5563-5567  (2009)]. For temperature $T$ below that of the onset to supercooled behavior, $T_{\mathrm{o}}$, the logarithm of structural relaxation time, $\log \tau$, 
is given by the parabolic form $\log (\tau/\tau_{\mathrm{o}}) \, =\, J^2( 1/T - 1/T_{\mathrm{o}})^{2}$, where $J$ and $\tau_{\mathrm{o}}$ are temperature independent.  This paper presents further applications of this formula.  In particular, it is shown that the effects of attractive forces in numerical simulation of glass forming liquids can be logically organized in terms of $J$ and $T_{\mathrm{o}}$. Further, analysis of experimental data for several systems suggests that $J$ and $T_{\mathrm{o}}$ are material properties.  In contrast, values of similar parameters for other fitting formulas are shown to depend not only upon the material but also upon the range of data used in fitting these formulas.  Expressions demonstrated to fail in this way include the Vogel-Fulcher-Tammann formula, a double-exponential formula, and a fractional exponential formula. \end{abstract}

\maketitle
\section{Introduction}

\begin{table*}[t!]
\begin{minipage}{\textwidth}
\caption{Comparison between VFT and parabolic fits for supercooled liquid OTP\\}
\begin{tabular}{ | c | c | c  | c | c || c  | c | c | c | }
\cline{2-9}
\multicolumn{1}{  c | }{} & \multicolumn{4}{ c || }{VFT\footnotemark[1]} & \multicolumn{4}{c | }{Parabolic\footnotemark[1]}\\
\hline

Fitted temperature range & $\log \eta^{(\infty)}_\mathrm{VFT}$/Poise & $A$/K$^{-1}$ & $T_{\mathrm{VFT}}$/K & Error\footnotemark[2]  &$J$/K$^{-1}$ & $T_{\mathrm{o}}$/K & $\log \eta_{\mathrm{o}}$/Poise  & Error\footnotemark[2]\\
\hline \hline
239 K - 350 K&  -8.98 & $1.03 \cdot 10^3$ & 191 &0.20  	& $2.79 \cdot 10^3$ & 352 & -1.88 & 0.064  \\
267 K - 350 K& -4.30 & 254 & 238 &  	51 & $2.82 \cdot 10^3$ & 351 & -1.86 & 0.087 \\
239 K - 265 K& -8.38 & 999 & 191 &  0.29	& $2.82 \cdot 10^3$ & 349 & -1.60 &  0.13 \\  \hline
\end{tabular}
\noindent \footnotetext[1]{  \ listed fitting parameters are determined by minimizing the mean square deviation between the fitting function and experiment for the specified range of data.}
\noindent \footnotetext[2]{ \ Root-mean-square deviation between fitted function and experiment over entire set of supercooled data.  That is, with the full set of data points, $\eta_i$ at the respective temperature $T_i$, $i = 1, 2, ..., N$, where $N$ is the total number of supercooled liquid data points determined by locating $T_{\mathrm{o}}$, Error$^2 = (1/N)\,\sum_{i=1}^{N} [\log \eta(T_i) \,-\,\log \eta_i]^2$, where $\eta(T)$ is the fitting function at temperature $T$ with parameters determined by minimizing the mean square deviation of the fitting function from experiment in the indicated range of experimental data.}

\label{tab:vftcomp}
\end{minipage}
\\
\begin{minipage}{\textwidth}
\caption{Comparison between double exponential and parabolic fit for supercooled liquid B$_{2}$O$_{3}$}
\begin{tabular}{ | c | c | c  | c | c || c  | c | c | c | }
\cline{2-9}
\multicolumn{1}{  c | }{} & \multicolumn{4}{ c || }{Double Exponential\footnotemark[1]} & \multicolumn{4}{c | }{Parabolic\footnotemark[1]}\\
\hline
Fitted temperature range & $\log \eta_{\mathrm{dx}}^{(\infty)}$/Poise & $K$ & $C$ & Error\footnotemark[2] & $J$/K$^{-1}$ & $T_{\mathrm{o}}$/K & $\log \eta_{\mathrm{o}}$/Poise &  Error\footnotemark[2] \\
\hline \hline
533 K - 970 K& 0.524 & 511 & $1.41 \cdot 10^3$ & 0.13	& $3.51 \cdot 10^3$ & $1.07\cdot 10^3$ & 2.96 & 0.087  \\
675 K - 820 K&  1.61& 169 & $2.01 \cdot 10^3$ & 	0.46	& $3.46 \cdot 10^3$ & $1.05\cdot 10^3$  & 3.14 & 0.20\\
533 K - 675 K& -0.940 & 974 & $1.12 \cdot 10^3$ &  	0.25	& $3.74 \cdot 10^3$ & $1.01 \cdot 10^3$ & 3.14  & 0.14\\ \hline
\end{tabular}
\noindent \footnotetext[1]{ \
Listed fitting parameters are determined by minimizing the mean square deviation between the fitting function and experiment for the specified range of data.}
\noindent \footnotetext[2]{ \
As defined in Table \ref{tab:vftcomp}.
}
\label{tab:decomp}
\end{minipage}
\\
\begin{minipage}{\textwidth}
\caption{Comparison of Fractional Exponential and Parabolic fit for PPG}
\begin{tabular}{ | c | c | c  | c | c || c  | c | c | c | }
\cline{2-9}
\multicolumn{1}{  c | }{} & \multicolumn{4}{ c || }{Fractional Exponential\footnotemark[1]} & \multicolumn{4}{c | }{Parabolic\footnotemark[1]}\\
\hline
Fitted temperature range& $\log \tau_{\mathrm{c}}^{(\infty)}$/s & $X$ & $T_{\mathrm{c}}$/K & Error\footnotemark[2] & $J$/K$^{-1}$ & $T_{\mathrm{o}}$/K & $\log \tau_{\mathrm{o}}$/s &  Error\footnotemark[2] \\
\hline \hline
200 K - 240 K	& -5.49	& 72.4 	& 244 	& 0.068 	& $2.26 \cdot 10^3$ 	& 264  	& -6.12 	& 0.042	\\
222 K - 240 K	& -6.66	& 41.8	& 265 	 & 0.33	& $2.27 \cdot 10^3$ 	& 265 	& -6.25 	& 0.13	\\
200 K -  218 K	& -3.73	& 116	& 227 	 & 0.11\footnotemark[3] 	& $2.29 \cdot 10^3$ 	& 265 	& -6.39 	& 0.088\footnotemark[3]$^{,}$\footnotemark[4]  \\ \hline
\end{tabular}
\noindent \footnotetext[1]{ \
Listed fitting parameters are determined by minimizing the mean square deviation between the fitting function and experiment for the specified range of data.}
\noindent \footnotetext[2]{ \
As defined in Table \ref{tab:vftcomp}.
}
\footnotetext[3]{\ When the fit range is restricted to the lowest temperatures, $T_\mathrm{c}$ is lower than the lowest data point available for PPG.  Therefore, the Error is calculated only for $T <$ 227 K for both the fractional exponential and the parabolic fit for sake of comparison.  This requires that the seven highest temperature points be excluded from consideration.  If those points are included, the Error for the fractional exponent over the entire range considered would become undefined.  }
\footnotetext[4]{\ Over the entire range for which $T < T_{\mathrm{o}} $ = 264 K, the Error for the the parabolic fit  using the parameters for this range becomes 0.13.}
\label{tab:fecomp}
\end{minipage}
\end{table*}

Transport properties of supercooled glass forming liquids are strong functions of temperature, increasing by typically ten or more orders of magnitude while absolute temperature is reduced by only ten or twenty percent~\cite{GlassReview, BinderKob}.  Most often this growth with respect to lowering temperature is super-Arrhenius and materials that behave in this way are called fragile. Seemingly varied behaviors of fragile glass formers are catalogued in terms of a property known as fragility~\cite{Fragility}.  This property measures the relative rate of change of a transport property as temperature is reduced near the glass transition temperature $T_{\mathrm{g}}$ -- the point at which the material falls out of equilibrium.  Fragility varies significantly from one material to another~\cite{VariedData}.  Yet, despite this variability, we have found that all such transport data can be collapsed to a non-singular function of temperature $T$~\cite{YSE_JP_DC},
\begin{equation}
\log \left(\tau /  \tau_\mathrm{o} \right)= J^2 \left( \frac{1}{T} - \frac{1}{T_\mathrm{o}} \right)^2\,,\,\,\,\,\,\,\,T<T_{\mathrm{o}}\,\,, 
\label{eqn:parab}
\end{equation}
where $\tau$ stands for the transport property, such as structural relaxation time or viscosity (which may also be denoted as $\eta$), and $\tau_{\mathrm{o}}$ refers to that same property at the onset temperature, $T_\mathrm{o}$.

The onset temperature marks the crossover from normal liquid behavior to supercooled liquid behavior.  Above that temperature, transport is unremarkable, indeed nearly temperature independent~\cite{JonasScience, WCA, BookHarris1980DiffuionInLiquids}.  It is only below that temperature where most equilibrium liquid transport properties depend sensitively upon temperature.  It is generally accepted that this crossover is due to the appearance of local rigidity or dynamical constraints, and this feature makes molecular motion rare and activated~\cite{DCJPGreview, JBJPBreview}. It is similarly accepted that correlations between motions in different regions of space cause activation barriers to grow with growing length scale, and this growth produces super-Arrhenius behavior for $T<T_{\mathrm{o}}$ (for example, Refs.~\cite{BinderKob, DCJPGreview, JBJPBreview, AdamGibbs, RFOT}).  But the mechanism and nature of this growth remain unresolved.  

To reach a resolution, it would be helpful to know if there is a precise and general form of temperature dependence for supercooled liquid transport.  From this perspective, it seems significant that throughout the range $T_{\mathrm{g}} < T < T_{\mathrm{o}}$\,, where transport properties vary by many orders of magnitude, all equilibrium transport of fragile glass formers is described by Eq. \ref{eqn:parab}.  This formula is an exact result for a class of dynamical models~\cite{sollich,DCJPPNAS,kcm,nef,fast}.  Can other formulas based on other models prove equally satisfactory?  We address this question in this paper by comparing Eq. \ref{eqn:parab} with a few other proposed relationships.  The alternatives are the commonly employed Vogel-Fulcher-Tammann (VFT) form \cite{VFTREF}, the recently advocated double exponential form~\cite{Mauro}, and a variant of Eq. \ref{eqn:parab} with a fractional exponent less than 2 \cite{fracExp}.  Each of these, like Eq. \ref{eqn:parab}, have three parameters.  (We exclude formulas with four~\cite{Kivelson}, or more, parameters from our consideration.) We show that the parameters of Eq. \ref{eqn:parab} are material properties in the sense that their values for a given system are reasonably invariant to the number and range of experimental data points. The parameters associated with the other considered forms do not have this quality.

The fact that data can be collapsed with Eq. \ref{eqn:parab} using parameters that are stable with respect to the range of data implies that this relationship can be used to predict results of measurements not yet performed.  It also provides a means to sensibly organize data.  Xu, Liu and Nagel  \cite{PeopleWhoCiteUs} applied a generalization of Eq. \ref{eqn:parab} in this way to collapse seemingly disparate data on a soft-sphere system~\cite{Berthier_soft}.  We further illustrate the organizational power in this paper by considering the role of attractive forces in supercooled transport.  Such forces can produce large effects at supercooled conditions~\cite{Berthier_WCALJ}.  We show that these effects reflect modest trends in the parameters $J$ and $T_{\mathrm{o}}$.

\section{Comparative Studies}
\subsection{VFT form compared with parabolic form}

\begin{figure}[t!]
\vspace{0.025in}
\includegraphics[width=3in]{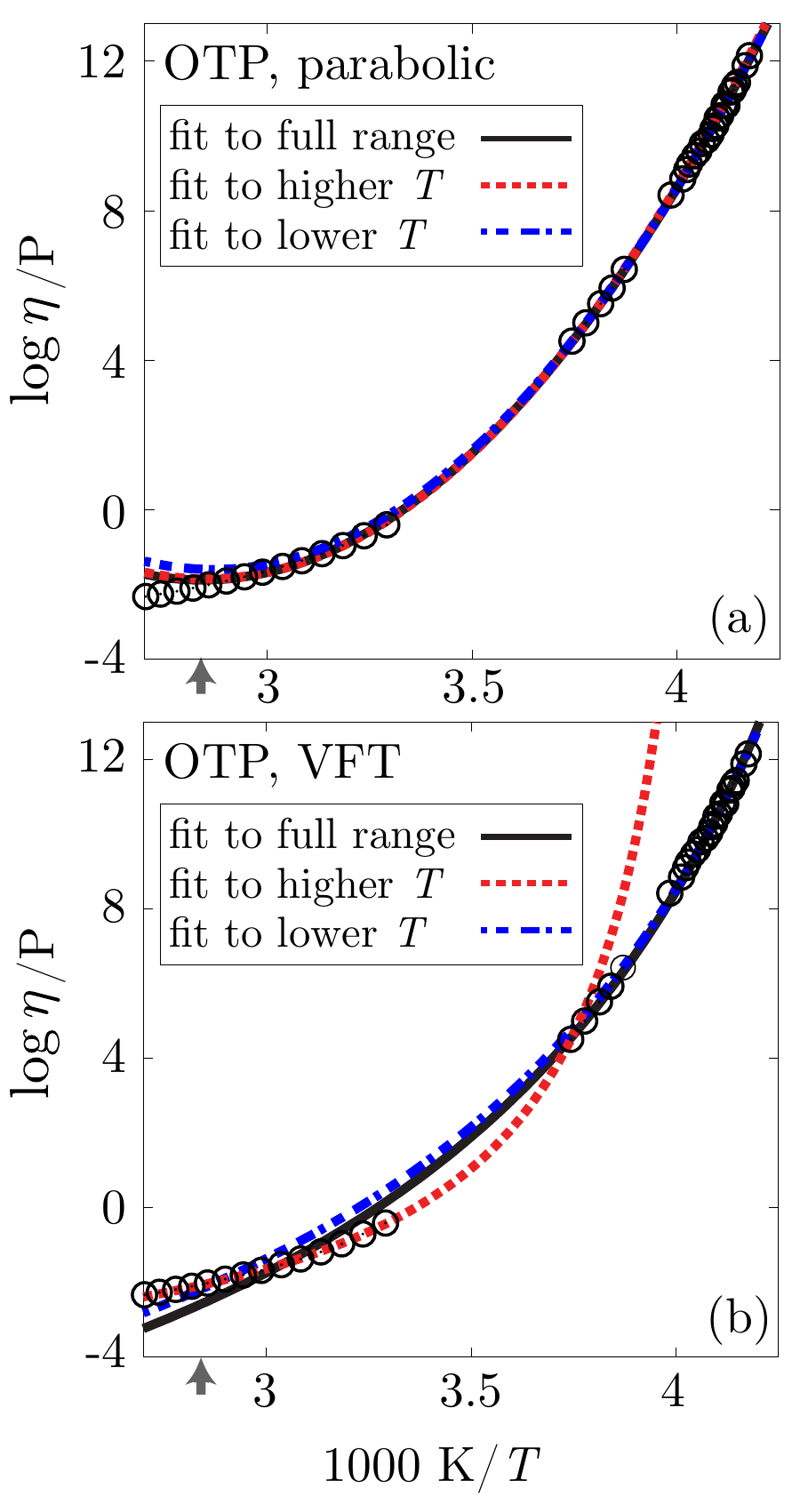}
\caption{\label{fig:vftcomp} Examples of using Eq. \ref{eqn:parab} (a) or Eq. \ref{eqn:vft} (b) to fit viscosity, $\eta$, of a supercooled liquid as a function of reciprocal temperature, $1/T$.  The circles are experimental data \cite{Mauro_OTP}.  Three fits are shown for both equations.  For one fit, parameters are determined by minimizing the mean square deviation between functional form and experiment for the full range of supercooled data, for the other two, parameters are found by minimizing the mean square deviation between the functional form and a subset of that data, the subset being either the higher temperature range of data or the lower temperature range of data.  See Table \ref{tab:vftcomp} for specified ranges and parameters.  The arrow indicates the value of $1/T_{\mathrm{o}}$, marking the crossover between normal and supercooled liquid behaviors.}
\end{figure}

\begin{figure}[t!]
\includegraphics[width=3in]{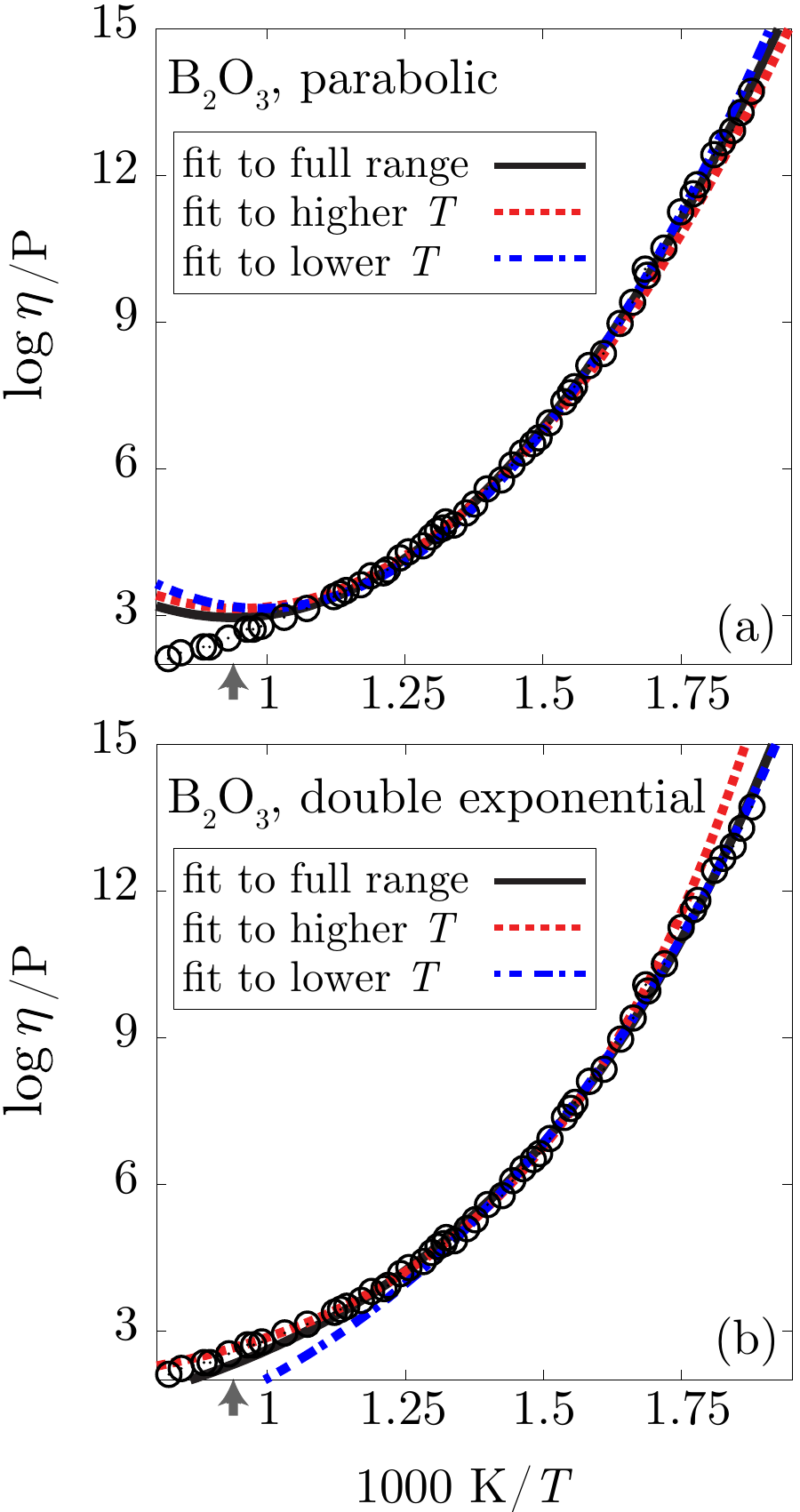}
\caption{\label{fig:decomp} Examples of using Eq. \ref{eqn:parab} (a) or Eq. \ref{eqn:doubleExp} (b) to fit viscosity, $\eta$, of a supercooled liquid as a function of reciprocal temperature, $1/T$.  The circles are experimental data \cite{B2O3}.  Three fits are shown for both equations.  For one fit, parameters are determined by minimizing the mean square deviation between functional form and experiment for the full range of supercooled data, for the other two, parameters are found by minimizing the mean square deviation between the functional form and a subset of that data, the subset being either the higher temperature range of data or the lower temperature range of data.  See Table \ref{tab:decomp} for specified ranges and parameters.  The arrow indicates the value of $1/T_{\mathrm{o}}$, marking the crossover between normal and supercooled liquid behaviors.} 
\end{figure}
The VFT formula for transport property $\tau$ is
\begin{equation}
\log(\tau /  \tau_{\mathrm{VFT}}^{(\infty)}) = \frac{A}{T-T_{\mathrm{VFT}}},
\label{eqn:vft}
\end{equation}
where $ \tau_{\mathrm{VFT}}^{(\infty)}$, \,$A$, and $T_{\mathrm{VFT}}$ are fitting parameters.  It is possibly the most common of all expressions used in glass physics, often referred to as a law \cite{Law}.   This formula is often invoked to attribute slowing dynamics in a supercooled liquids to a thermodynamic transition at the temperature $T_{\mathrm{VFT}}$.  Adam-Gibbs theory \cite{AdamGibbs} and random-first-order theory \cite{RFOT} connect this transition to an entropy crisis, where $T_{\mathrm{VFT}}$ is a recorder of a Kauzmann temperature.  The latter, $T_{\mathrm{K}}$, is a temperature at which the entropy of the supercooled liquid equals that of the ordered solid.  (A formula analogous to Eq. \ref{eqn:vft}
is an exact result for a class of stochastic models~\cite{Toninelli06}.)

It is often asserted that $T_{\mathrm{VFT}} \approx T_{\mathrm{K}}$ is a good approximation \cite{RFOT}, but it has been observed that experimental evidence is mixed~\cite{Tanaka}.  To analyze the issue quantitatively, note that both $T_{\mathrm{VFT}}$ and $T_{\mathrm{K}}$ are constrained to lie below the glass transition temperature, $T_{\mathrm{g}}$, and the relative difference from that reference temperature is the pertinent measures.   According to the tabulated temperatures for 33 different liquids \cite{Angell_Table}, the quantities $\left(T_{\mathrm{g}}/T_{\mathrm{VFT}}-1 \right)$ and $\left(T_{\mathrm{g}}/T_{\mathrm{K}}-1 \right)$ are typically a few tenths.  Their ratio, $R = \left(T_{\mathrm{g}}/T_{\mathrm{VFT}}-1 \right)/ \left(T_{\mathrm{g}}/T_{\mathrm{K}}-1 \right)$, often differs significantly from unity.  In particular, averaging over the 33 different liquids yields a mean value of $\langle R \rangle = 1.15$ and a root-mean-square deviation $ \langle \left( R-\langle R \rangle \right)^2 \rangle^{1/2}= 0.613$.  In other words, it is just as likely that $T_{\mathrm{VFT}}$  and  $T_{\mathrm{K}}$ will differ significantly as it is that they will be similar.

Independent of an alleged connection to thermodynamics, the utility of the VFT formula can be tested by examining whether its parameters are material properties.  We do so for one typical fragile liquid, $o$-terphenyl (OTP).  What we illustrate with this system is representative of what we generally find for a wide selection of supercooled liquids.  

The onset temperature for OTP is about 350 K, and the glass transition temperature is 239 K.  Data exists over this entire range~\cite{Mauro_OTP}.  Our earlier paper~\cite{YSE_JP_DC} considered the lower half of this range, but here we consider the full supercooled range for the purpose of exploring sensitivity of parameters.  (We omit the normal liquid range $T > T_{\mathrm{o}}$, and only remark that the deficiencies of VFT we are about to demonstrate are more severe when the temperature range is extended to include the normal liquid regime.)  
With this range of data, we have fit the parameters for the VFT formula and those for the parabolic formula in three different ways: by including data from only the range of higher supercooled temperatures ($267 \text{ K}<T<350 \text{ K}$), by including data from only the range of lower supercooled temperatures ($239 \text{ K} <T< 267 \text{ K}$), and by including data from the entire range of supercooled temperatures ($ 239 \text{ K}<T<350 \text{ K}$).  The fitting parameters obtained in these ways are given in Table~\ref{tab:vftcomp}.

\begin{figure}[t!]
\includegraphics[width=3in]{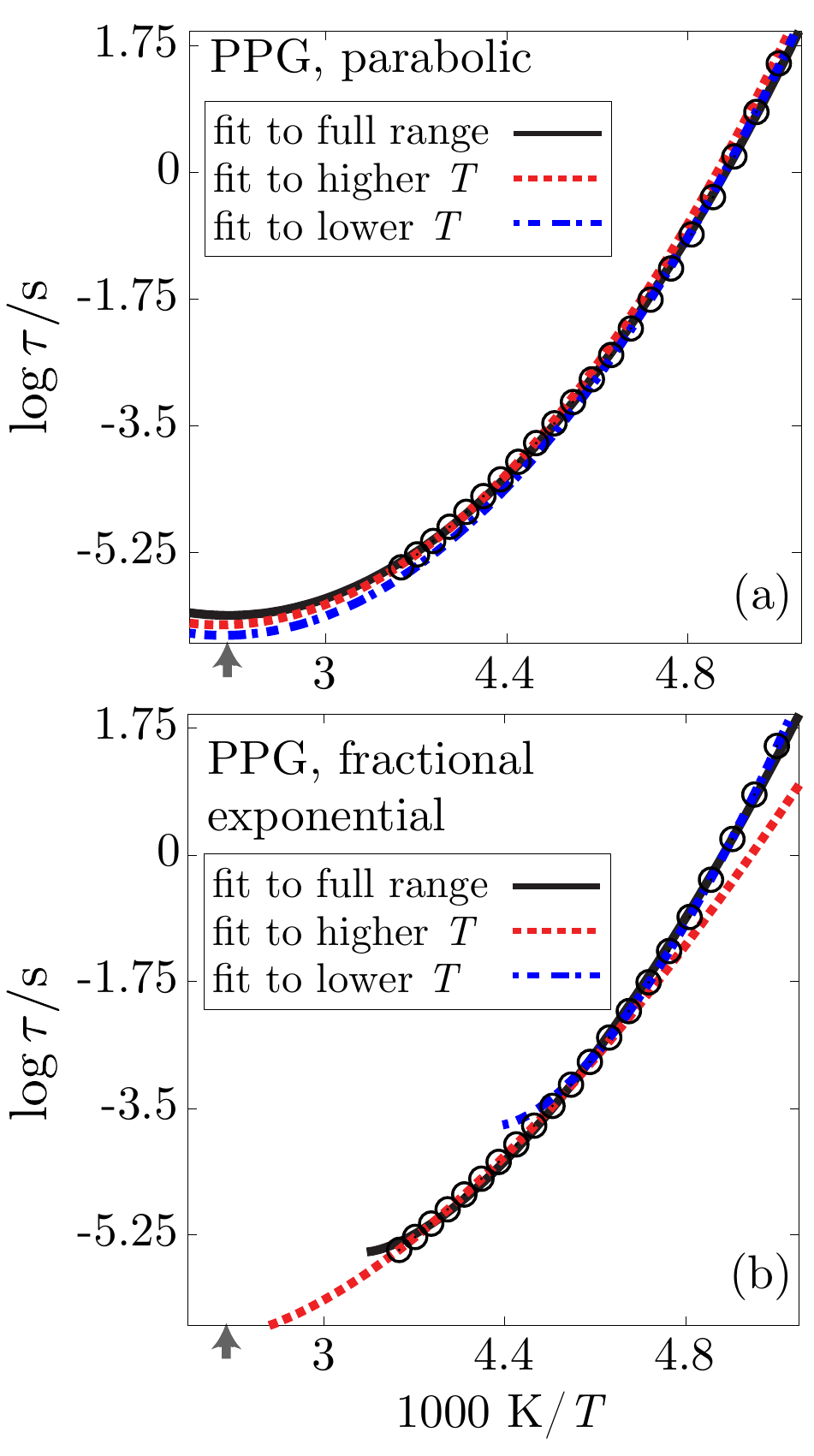}
\caption{\label{fig:fecomp}Examples of using Eq. \ref{eqn:parab} (a) or Eq. \ref{eqn:fracExp} (b) to fit relaxation time, $\tau$, of a supercooled liquid as a function of reciprocal temperature, $1/T$.  The circles are experimental data \cite{Dyre}.  Three fits are shown for both equations.  For one fit, parameters are determined by minimizing the mean square deviation between functional form and experiment for the full range of supercooled data, for the other two, parameters are found by minimizing the mean square deviation between the functional form and a subset of that data, the subset being either the higher temperature range of data or the lower temperature range of data.  See Table \ref{tab:fecomp} for specified ranges and parameters.  The arrow indicates the value of $1/T_{\mathrm{o}}$, marking the crossover between normal and supercooled liquid behaviors. The leftmost endpoints of the fit lines in figure (b) indicate the upper-temperature end point for applying Eq. \ref{eqn:fracExp}.  This temperature changes depending upon the range of data considered, and data for $T>T_{\mathrm{c}}$ must be excluded from fits using Eq. \ref{eqn:fracExp}.} 
\end{figure}

The entries to this table show a significant difference between parameters found from fitting the full range of data and those found from fitting only the higher temperature data.  The VFT parameters found from fitting only the lower temperature data agree well with those found from fitting all supercooled data at the expense of a larger fitting error.  For these cases, the error in reproducing the full range of supercooled liquid data is reasonably small, but at the expense of an unphysical value for the reference viscosity, $\eta_{\mathrm{VFT}}^{(\infty)}$.  Figure \ref{fig:vftcomp} shows poor agreement between experiment and the VFT formula for high temperature data when only low temperature data is included in fitting (and vice versa).  In the latter case, when only the higher temperature range of supercooled data is used to fit VFT parameters, the value of $T_{\mathrm{VFT}}$ is close to $T_{\mathrm{g}}$, causing the huge error reported in the second row of Table~\ref{tab:vftcomp}.  In contrast, the parameters and quality of fits found with the parabolic form change little between the full set of data or either of the subsets of data.  

Thus, parameters of the parabolic form appear to represent properties of the system, and those determined over one range of temperatures can be used to reliably predict properties over another range of temperatures.  The parameters of the VFT form, however, depend upon both the properties of the system and the range of data considered, and as such the VFT form cannot predict properties outside the range over which it has been fit.

\subsection{Double-exponential form compared with parabolic form}
The double-exponential formula for transport property $\tau$ is
\begin{equation}
\log(\tau /  \tau_{\mathrm{dx}}^{(\infty)}) = (K/T)\exp(C/T)
\label{eqn:doubleExp}
\end{equation}
where $ \tau_{\mathrm{dx}}^{(\infty)}$, \,$K$, and $C$ are fitting parameters.  A special case, where $K$ and $C$ are of the same order, is the behavior of two-spin (or two-particle) facilitated lattice models \cite{Toninelli05}.  Over the lower half of the temperature range between $T_{\mathrm{o}}$ and $T_{\mathrm{g}}$, the double exponential has been used to successfully collapse transport data~\cite{Rossler}.  Mauro and co-workers~\cite{Mauro} and others~\cite{OtherDoubleExpPaper} have applied it to a broader range and report that Eq. \ref{eqn:doubleExp} is superior to both the VFT form, Eq. \ref{eqn:vft}, and the parabolic form, Eq. \ref{eqn:parab}.  In fact, the double-exponential form suffers from the same malady as the VFT form when applied to fit data over the full range of supercooled temperatures.  As a data set is enlarged, its parameters fail to converge, implying these parameters are not material properties and the formula cannot be used to predict data not yet measured.   The problem is not as severe as it is for the VFT form, owing to the fact that the double exponential is not singular while the VFT expression is singular.  But the deficiency of the double-exponential form is nonetheless significant, as illustrated in Figure \ref{fig:decomp} and Table \ref{tab:decomp}. 

Our illustration considers the inorganic glass forming liquid B$_{2}$O$_{3}$ \cite{B2O3}.  The behaviors found for this system are typical of what we find for several other systems.  For this particular system, the onset temperature for this liquid is close to 1000K, and the glass transition temperature for this liquid is about 500K.  Fitting data over this entire range, the double-exponential form proves reasonably accurate, but its parameters change markedly as the range of fitted data changes.  Fitting only higher temperature data, $ 675\text{ K} \leqslant T \leqslant 820 \text{ K}$ , yields a function that inaccurately describes the lower temperature data, and fitting only lower temperature data, $ 533 \text{ K} \leqslant T \leqslant 675 \text{ K}$, yields a function that inaccurately describes the higher temperature data.  In contrast, the parameters and excellent quality fits of the parabolic form change little as the range of fitted supercooled data changes. 

Thus, for the purpose of employing a particular functional form to predict low temperature properties from measured properties at higher supercooled temperatures (or vice versa), Eq. \ref{eqn:doubleExp} is superior to the VFT expression, but this double exponential form advocated by Mauro and coworkers~\cite{Mauro} is inferior to the parabolic form, Eq. \ref{eqn:parab}. 

\begin{figure*}[t!]
\begin{center}
\includegraphics[width=6.5in]{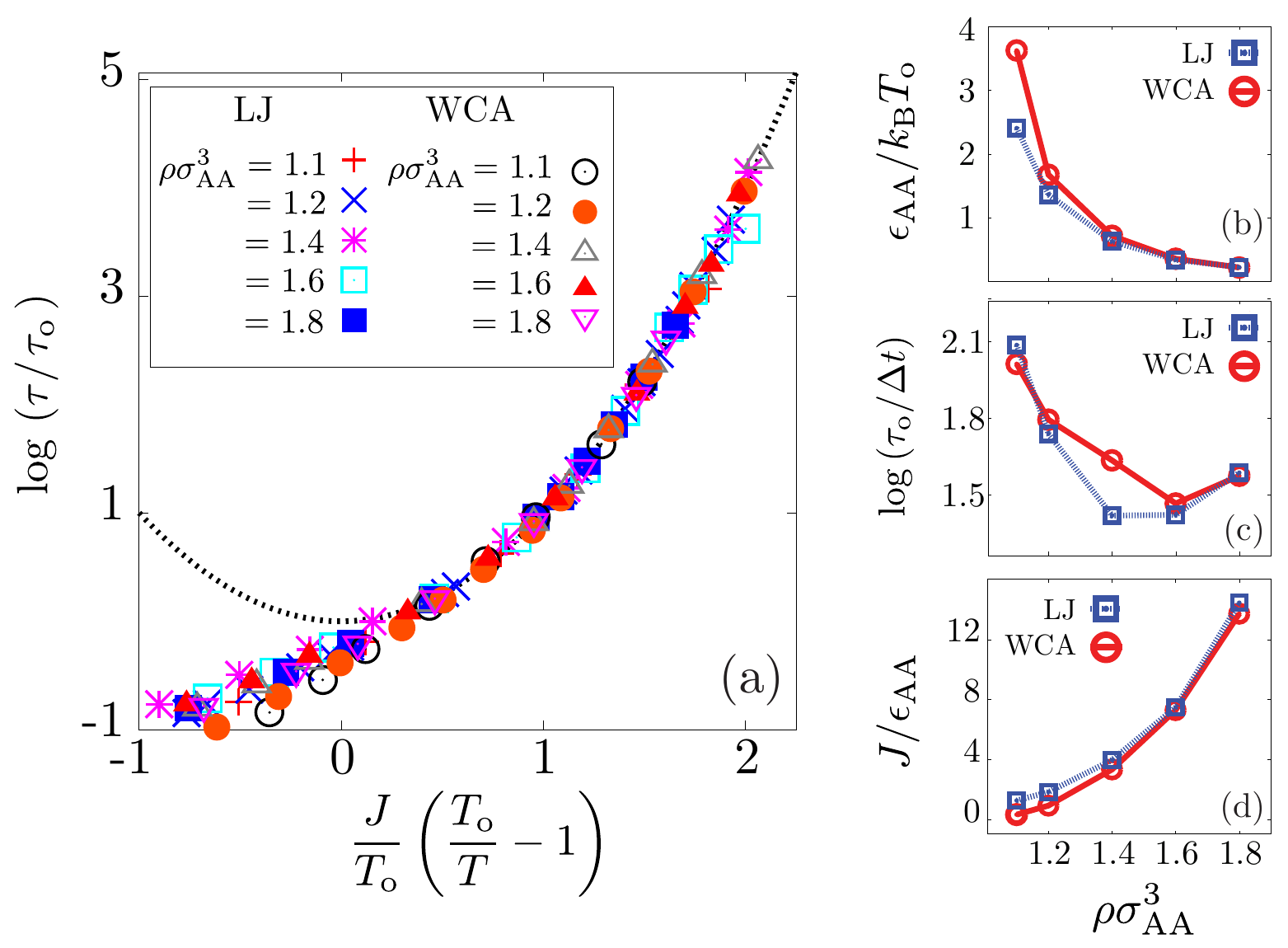}
\end{center}
\caption{\label{fig:wcalj} 
(a) parabolic collapse for WCA \& LJ binary mixture simulation data for various densities, $\rho$ from \cite{Berthier_WCALJ}.  Parameters of fit are described in Eq. \ref{eqn:parab}. Fitting parameter trends using Eq. \ref{eqn:parab} for the Kob-Andersen LJ and corresponding WCA binary mixtures from \cite{Berthier_WCALJ} for various net particle densities,  $\rho$.  (b) The inverse onset temperature, $1/T_\mathrm{o}$, as a function of density  $\rho$.  (c) Logarithm of relaxation time at the onset temperature, $\log \tau_\mathrm{o}$, as a function of $\rho$.  (d) Transport energy parameter, $J$, as a function of  $\rho$.  Dotted line in (a) is the universal parabolic form.  Lines connecting points in (b), (c) and (d) are guides to the eye.  Error estimates are the size of the symbols.  The unit of time is $\Delta t = (m \sigma^2_\mathrm{AA}/48 \epsilon_\mathrm{AA})^{1/2}$. }
\end{figure*}

\begin{figure}[t!]
\includegraphics[width=3.4in]{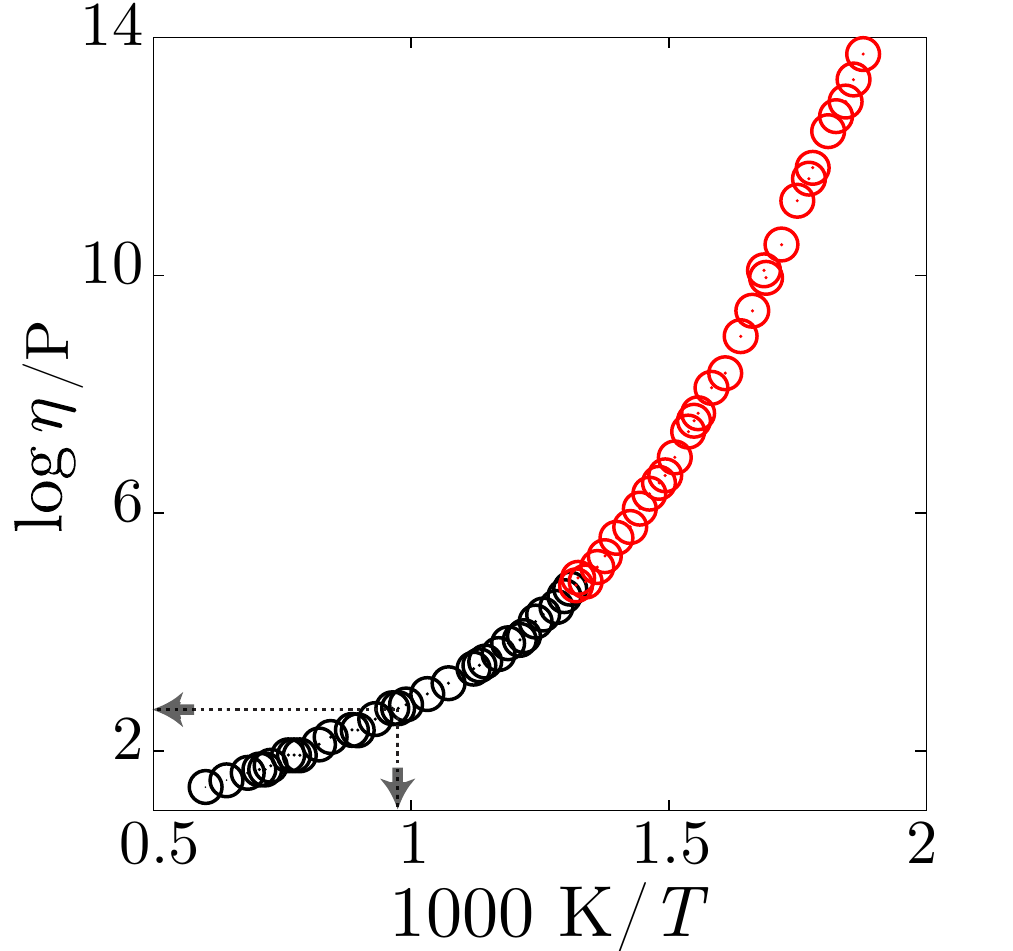}
\caption{\label{fig:b2o3min} 
An example of fitting parameter convergence for B$_{2}$O$_{3}$.  Red points indicate the minimal data set required to converge fitting parameters when beginning from the low temperature data and extrapolating downward.  Arrows indicate the location of the onset of fragile behavior. } 
\end{figure}

\subsection{Fractional-exponent form compared with parabolic form}
In both the VFT and double-exponential forms, temperature variations are more rapid than in the parabolic form.  What if temperature variation is taken to be less rapid?  An alternative of this type is the specific fractional-exponent form suggested by numerical solutions to Saltzmann and Schweizer's theory for structural relaxation in polymer melts \cite{fracExp},
\begin{equation}
\log(\tau /  \tau_{\mathrm{c}}) = X\left(T_{\mathrm{c}}/T - 1 \right)^{1.57}\,.
\label{eqn:fracExp}
\end{equation}
We illustrate the performance of this expression in Figure \ref{fig:fecomp} and Table \ref{tab:fecomp}.  We do so with data for the glass forming polymer melt polypropylene glycol (PPG) \cite{Dyre}.  The onset temperature for this liquid is about 264 K, and its glass temperature is about 199 K \cite{YSE_JP_DC}.  The data covers most of this supercooled region, but not all the way up to the onset.  As in the previous two case studies, we consider three ranges of the existing data:  the entire data set, which extends up to about 10\% of the onset temperature, $T_\mathrm{o}$, a lower temperature subset of that data, and a higher temperature subset of that data.  As with the VFT and double-exponential forms, we find that the fitting parameters for this fractional-exponential form depend upon the range of data considered.  Our illustration of this fact is typical of what we find when treating other materials with the same analysis. 

Due to the nature of the fractional exponent, this functional form can be used to fit data for $T > T_\mathrm{c}$.  If the exponent were 2, rather than 1.57, this temperature would be the onset temperature.  When using an exponent of 1.57 with subsets of the data where temperatures are all much smaller than $T_{\mathrm{o}}$ often one obtains $T_{\mathrm{c}} < T_{\mathrm{o}}$, so that less than the complete set of supercooled data can be covered.  Further, unlike the parabolic form with parameters that vary little with changing data sets, the best fits of Eq. \ref{eqn:fracExp}  produce parameters $T_{\mathrm{c}}$ and $X$ that vary widely from one subset of data to another. Moreover, the function obtained fitting parameters to the higher temperature range of data, $ 222\text{ K} \leqslant T \leqslant 240\text{ K}$, provides an inaccurate representation of the data at the lower temperatures, and the function obtained fitting parameters to the lower temperature range of data, $200\text{ K} \leqslant T \leqslant  218\text{ K}$, provides an inaccurate representation of the data at the higher temperatures.  In contrast, the excellent quality of fits obtained with the parabolic form, Eq. 1, change little as the range of fitted data change.

Thus, the fractional exponent 1.57 proves to be less satisfactory than that of the parabolic form, Eq. \ref{eqn:parab}.

\section{Role of Attractive Forces}

The prior section presents evidence in support of Eq.\ref{eqn:parab} as a universal form for transport properties of fragile supercooled liquids. In this section, we show that by accepting the validity of this form we sensibly organize recent simulation results that might appear puzzling in the absence of this organization.  In particular, Berthier and Tarjus\cite{Berthier_WCALJ} have shown that at some supercooled temperatures and densities relaxation in the Kob-Andersen Lennard-Jones (LJ) mixture\cite{KobAnd} is orders of magnitude slower than that in the corresponding Weeks-Chandler-Andersen\cite{WCA1971} (WCA) mixture. The difference between the WCA and the LJ potential reflects the significance of attractive forces, which are present in the LJ mixture and absent in the WCA mixture. Does the Berthier-Tarjus finding reveal a new mechanism for glassy physics, one that does not follow from constraints and local rigidity imposed by repulsive forces? 

The effects of attractive interactions uncovered by Berthier and Tarjus appear in dynamics, but not structure. They show that the pair distribution functions of the two mixtures differ very little. Dynamics is associated with fluctuations away from mean local structure. Repulsive forces dominate the most likely equilibrium arrangements.  They also constrain motions of particles, so that any spatial reorganization is rare and requires the coordinated displacements of several particles. As such, potential energy barriers to reorganization will have contributions from the interactions between several pairs of particles.  The sums of small contributions from many particles can become significant.  In other words, it can be due to the local rigidity from packing forces that small effects from attractive forces can become notable.  

To pursue this idea, we have examined the effects of attractive forces on supercooled fluid transport through the behaviors of $J$ and $\To$ because these parameters characterize the energetics of these collective displacements.\cite{DCJPPNAS}  We have fit the simulation data\cite{Berthier_WCALJ} reported to determine the values of these parameters as functions of the liquid-mixture density. These results are shown in \ref{fig:wcalj} along with the reference time scale $\tau_{\mathrm{o}}$. While the values of $J$, $\To$, and $\log \tau_\mathrm{o}$ do vary slightly between the two types of mixtures at the same density, their overall trends as functions of density are the same.  Both liquid mixtures therefore seem to behave as similar supercooled liquids.

Berthier and Tarjus remark on the ``enormous'' factor of $10^{3.2}$ by which $\tau$ for the LJ mixture differs from that for the WCA mixture at $\rho \sigma_{\mathrm{AA}}^{3} = 1.2$ and $k_B T / \epsilon_{\mathrm{AA}} = 0.41$. This effect reflects modest differences in $J/\epsilon_{\mathrm{AA}}$ and $k_B T_{\mathrm{o}} / \epsilon_{\mathrm{AA}}$, which are 1.8  and 0.73 for the LJ mixture, and 0.92 and 0.60 for the WCA mixture, respectively.  The onsets to supercooled behavior thus appear at similar temperatures, and the activation energies for dynamics differ by less than one intermolecular attractive energy. Moreover, as the density increases the characteristic parameters for the two systems seem to converge to the same respective values. These findings are consistent with empirically established fact that the basic principles of supercooled liquids are well captured by models without attractive intermolecular forces \cite{Ulf1,Ulf2}. 

Thus, while enhanced by the results of Berthier and Tarjus, the picture of glassy physics as a class of phenomena caused by local constraints and rigidity remains unchanged.

\begin{acknowledgments}
We thank L. Berthier for providing tables of numerical data, J. Mauro and Y. Yue for helpful correspondence, and U. Pedersen for useful discussions.  While carrying out this work, Y.S.E. was supported by NSF GRFP and ONL NDSEG fellowships, and D.C. was supported in its initial stages by the NSF and in its final stages by DOE Contract No. DE-AC02-05CH11231.
\end{acknowledgments}

\begin{figure*}[t!]
\begin{center}
\includegraphics[width=6.5in]{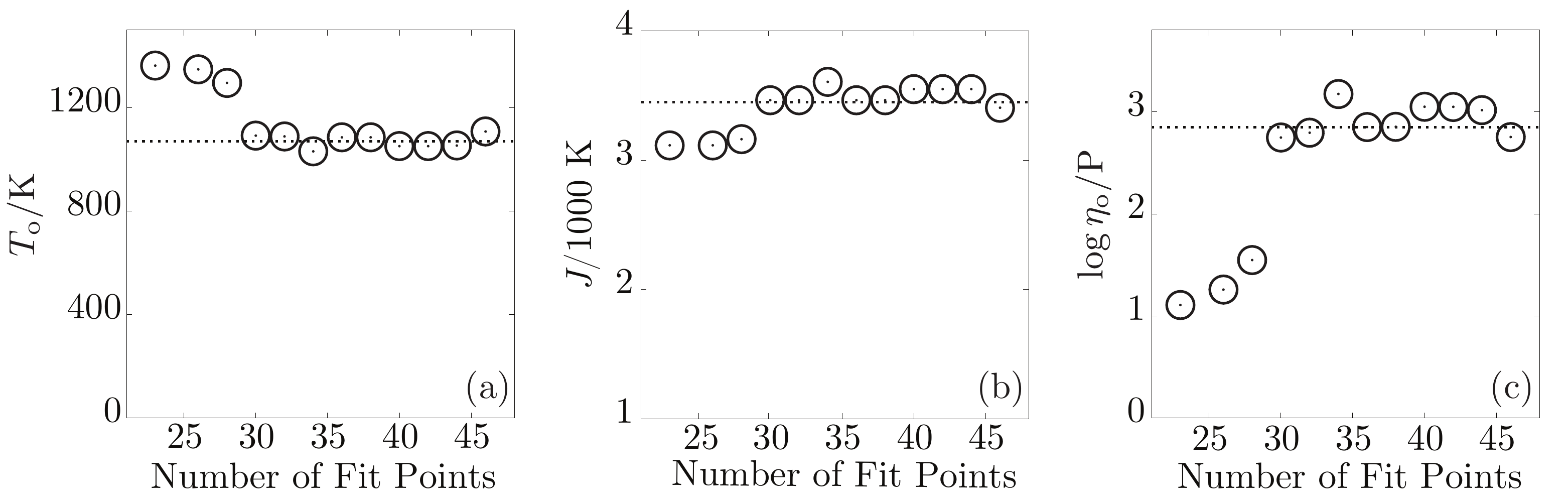}
\end{center}
\caption{\label{fig:converge} 
An example of fitting parameter convergence for B$_{2}$O$_{3}$.  In the fragile regime, 47 data points ranging continuously over 11 orders of magnitude in $\eta$ are available for fitting.  Including the liquid regime, there are 62 data points in this set spanning 13 orders of magnitude in $\eta$.  Initially, only the lowest temperature points are considered and subsequent fits with more points come from adding higher temperature points to the considered set. (a) shows the convergence of $T_\mathrm{o}$ as a function of number of points used for fitting.  (b) shows the convergence of $J$ as a function of number of points used for fitting.  (c) shows the convergence of $\eta_\mathrm{o}$ (analogous to $\tau_\mathrm{o}$) as a function of number of points used for fitting.  For this system, $T_\mathrm{o}=$ 1066 K, $J/T_\mathrm{o}=$ 3.3, and $\log \eta_\mathrm{o}/\mathrm{Poise} = 2.9$. } 
\end{figure*}

\appendix*
\section{How to fit data to the parabolic form.}

Here, we summarize the procedure we employ to fit data with the parabolic form.

\noindent 1. Relaxation time or viscosity data spanning several orders of magnitude are required.

\noindent 2. Examine the data to attempt to identify normal liquid and fragile regimes, and thereby obtain an approximate location of the onset lying between the two.

\noindent 3.  Fit data spanning several orders of magnitude starting with the lowest temperatures available, using a least squares analysis to obtain a first estimate of $J$, $\To$ and $\log \tau_{\mathrm{o}}$.
 
\noindent 4.  Use this estimate for $\To$ as an indication of the highest temperature point for which to fit the data and continue to add higher temperature points to your fit until the values of $J$, $\To$ and $\log \tau_\mathrm{o}$ have converged.  (An illustration of the minimal amount of data needed to fit viscosity data for B$_{2}$O$_{3}$ \cite{B2O3} is shown in Figure \ref{fig:b2o3min}.  An illustration of the parameter convergence for the B$_2$O$_3$ fit is shown in Figure \ref{fig:converge}.)  

As data points are added after convergence, small systematic changes in parameter values can occur.  Generally, $J$ and $T_\mathrm{o}$ tend to rise together where as $\log \eta_\mathrm{o}$ decreases and vice versa.  These variations reflect the presence of local minima in fits of nearly equal standard deviation that swap from being local to global, and their sizes appear to be negligible in comparison with statistical errors in experimental data. \\

\noindent Caveats:

\noindent A. While in our B$_{2}$O$_{3}$ example we have chosen to start at the lowest temperature point and add higher and higher temperature points until the parameters converge, it should be noted that often the lowest temperature points are the hardest to measure and therefore sometimes data in this region is not always as reliable as data taken from a more moderate, yet still supercooled regime.  It may be useful to check if a fit through such a moderate range properly predicts the lowest temperature data as an estimate for the quality of the lowest temperate data.\\ 
B. The idea that fragile glass former transport data may, at low temperatures, crossover again from fragile to Arrhenius has been proposed~\cite{DCJPPNAS} and recently considered further~\cite{Mallamace}.  At this moment, there is limited evidence to support or refute this claim.  Certain systems seems to exhibit this phenomenon \cite{Mallamace, WaterCrossover, Laughlin_1972}.  However, the systems considered are either confined to nano-scale pores, which may or may not reflect the behavior of macroscopic materials, or more recent experiments contradict those that exhibit a crossover \cite{SalolOTPCounterExamples}.  In the future, it may be necessary to reconsider this idea if lower temperature data can be collected for experimental systems.  In this case, for the parabolic fitting, extra caution should be taken with the data and a low temperature fragile-strong crossover point $T_\mathrm{x}$ should be identified.  Under these circumstances, fit the lowest temperature points with: 
$$
\log(\tau / \tau_{\mathrm{x}}) = E \left(\frac{1}{T} - \frac{1}{T_{\mathrm{x}}} \right)
$$
From this, choose a reasonable low-temperature end to the fragile fitting regime, $T_\mathrm{x}$, where the Arrhenius behavior at low temperatures diverges from the experimental or numerical data.

\newpage

\bibliographystyle{apsrev}


\end{document}